\documentclass
[preprint,prd,onecolumn,showpacs,showkeys,nobibnotes,nofootinbib,titlepage]{revtex4}%
\usepackage{amsfonts}
\usepackage{amsmath}
\usepackage{amssymb}
\usepackage{graphicx}
\usepackage{float}
\usepackage{rotating}%
\providecommand{\U}[1]{\protect\rule{.1in}{.1in}}
\pdfoutput=1

\ifx\pdfoutput\relax\let\pdfoutput=\undefined\fi
\newcount\msipdfoutput
\ifx\pdfoutput\undefined\else
\ifcase\pdfoutput\else
\ifx\paperwidth\undefined\else
\ifdim\paperheight=0pt\relax\else\pdfpageheight\paperheight\fi
\ifdim\paperwidth=0pt\relax\else\pdfpagewidth\paperwidth\fi
\fi\fi\fi
\begin{document}
\preprint{ }
\title[Black hole shadows]{Black hole shadows in fourth-order conformal Weyl gravity}
\author{Jonas R. Mureika\thanks{Email: jmureika@lmu.edu} and Gabriele U.
Varieschi\thanks{Email: gvarieschi@lmu.edu}}
\affiliation{Department of Physics, Loyola Marymount University - Los Angeles, CA 90045, USA}
\eid{ }
\eid{ }
\author{}
\affiliation{}

\begin{abstract}
We calculate the characteristics of the \textquotedblleft black hole
shadow\textquotedblright\ for a rotating, neutral black hole in fourth-order
conformal Weyl gravity. It is shown that the morphology is not significantly
affected by the underlying framework, except for very large masses. Conformal
gravity black hole shadows would also significantly differ from their general
relativistic counterparts if the values of the main conformal gravity
parameters, $\gamma$ and $\kappa$, were increased by several orders of
magnitude. Such increased values for $\gamma$ and $\kappa$ are currently ruled
out by gravitational phenomenology. Therefore, it is unlikely that these
differences in black hole shadows will be detected in future observations,
carried out by the Event Horizon Telescope or other such experiments.

\end{abstract}
\startpage{1}
\endpage{ }
\maketitle
\tableofcontents


\vskip 2cm

\section{\label{sect:introduction}Introduction}

Immediately following the centenary of Einstein's general relativity (GR), we
have been witness to a major test of the theory's foundational predictions.
Two separate detections by LIGO of gravitational waves from binary black hole
(BH) mergers \cite{Abbott:2016blz,Abbott:2016nmj,TheLIGOScientific:2016pea}
have provided incontrovertible experimental evidence of this long-predicted
feature of GR. A second test -- the imaging of a black hole's
\textquotedblleft shadow\textquotedblright/photosphere, and by proxy its event
horizon -- is looming near. Pioneered by the Event Horizon Telescope (EHT) and
BlackHoleCam consortia \cite{eht1,bhcam1}, it will involve targeting the
putative supermassive black hole Sagittarius A* (Sgr A*) at the center of the
Milky Way \cite{sgr}, as well as active galactic nuclei \cite{eht2}, and it is
anticipated that this will provide a crucial test of GR against competing
theories by allowing precision measurements of the horizon size
\cite{Johannsen:2015hib}.

This and other morphological characteristics of the Sgr A* shadow will allow
for precision measurement of the object's mass, but in principle can also be
used to probe the curvature, and thus the underlying gravitational theory.
Since astrophysical black holes are generally expected to be rotating and
neutral, a baseline standard for general relativity can be extracted from an
analysis of the Kerr solution \cite{Li:2013jra,Johannsen:2015qca}. Given this
fact, it is possible that such data will also be sensitive to modifications of
the underlying gravitational theory.

To date, a number of studies have addressed aspects of the shadow morphology
associated with alternate theories and extensions to general relativity, and
the literature is growing increasingly comprehensive. Select approaches
include analyzing the shadow characteristics of Kerr black holes with scalar
hair \cite{shadow8}, a five-dimensional Myers-Perry black hole
\cite{Papnoi:2014aaa}, distorted Schwarzschild \cite{shadow6} and Kerr black
holes \cite{shadow11}, as well as the possibility of observing double images
from a single black hole \cite{shadow12}. Additional shadow traits in other
extensions and alternatives to general relativity include noncommutative
gravity inspired black holes \cite{Wei:2015dua}, Modified Gravity
\cite{shadow7}, $f(r)$ gravity \cite{shadow9}, a rotating Einstein-Born-Infeld
black hole \cite{Atamurotov:2015xfa}, regular black holes \cite{shadow10}, and
rotating non-singular black holes \cite{Amir:2016cen}.

Furthermore, the EHT and related experiments could provide a novel test of
Hawking radiation and string inspired theories through near horizon effects.
Such phenomena will implicitly depend on the transfer of information across
the horizon, and could effectively address the information paradox. One such
example concerns the phenomenological impact of couplings between a black
hole's internal states and those immediately outside the horizon, which
manifest themselves as (quantum) spacetime fluctuations
\cite{giddings1,giddings2,giddings3,giddings4,giddings5}. Although appearing
at low energy scales, these fluctuations can be significant in magnitude, and
can deflect near-horizon geodesics that span distances on the order of the
black hole's radius. That is, the EHT can be used as an effective test of the
information paradox, and a coherent analysis of near-horizon physics may open
a window to quantum gravity.

The associated conclusions are mixed as to whether or not the modification of
choice will have any measurable impact on the shadow within the limit of the
experimental sensitivity. These range from shadow sizes both smaller and
larger than the general relativistic predictions, as well as morphological
discrepancies (\textit{e.g.} asymmetric shapes). As illustrative examples, the
Einstein-Born-Infeld shadow considered in \cite{Atamurotov:2015xfa} was found
to be smaller than that of a Reissner-Nordstr\"{o}m black hole. Other models
such as the Kerr black hole with scalar hair predict shadows of
distinguishable shape from those in general relativity \cite{shadow11}.
Differentiability of shadow characteristic by the EHT between Randall-Sundrum
and Einstein-Gauss-Bonnet black holes have also been considered
\cite{Johannsen:2015hib}. Perhaps the most compelling result is that for the
Modified Gravity black hole \cite{shadow7}, which predicts a shadow radius
bigger than that of the standard Kerr solution, depending on the
(non-universal) size of the Modified Gravity parameter $\alpha$.

In the following paper, we address aspects of the black hole shadow in a
fourth-order conformal Weyl gravity framework. Conformal gravity -- or CG for
short -- is a natural extension of Einstein's general relativity, originally
introduced by H. Weyl \cite{Weyl:1918aa}, revisited by P. Mannheim and others
\cite{Mannheim:1988dj,Kazanas:1988qa,Mannheim:2005bfa,Varieschi:2008fc,Varieschi:2008va,Varieschi:2014ata,Varieschi:2014pca}%
, and even considered by G. 't Hooft
\cite{Hooft:2014daa,Hooft:2010ac,Hooft:2010nc,tHooft:2011aa} as a possible key
towards a complete understanding of physics at the Planck scale.

Conformal gravity naturally addresses and solves several cosmological and
astrophysical problems \cite{Mannheim:2011ds}, such as the cosmological
constant problem \cite{Mannheim:1989jh,Mannheim:1999nc}, zero-point energy and
quantum gravity problems \cite{Mannheim:2009qi}, the fitting of galactic
rotation curves \cite{Mannheim:2010xw} and accelerating universe supernovae
data \cite{Mannheim:2005bfa}, without any dark matter or dark energy.

It was recently noted by Maldacena that there is a holographic connection of
CG to Einstein gravity (EG) \cite{Maldacena:2011mk}, in the sense that CG
reduces to EG for certain boundary conditions. Furthermore, it was shown by
Grumiller, Irakleidou, Lovrekovic, and McNees that by generalizing the
Starobinski boundary conditions in \cite{Maldacena:2011mk}, the aforementioned
CG solutions of Mannheim \textit{et al.} can reproduce the solutions of
Mannheim and Kazanas \cite{Mannheim:1988dj} and Riegert \cite{Riegert:1984zz}
with finite holographic response functions.

Although fourth-order gravity theories have long been thought to possess
ghosts when quantized, it has recently been suggested that CG may be
ghost-free \cite{Bender:2007wu,Mannheim:2007ki,Mannheim:2006rd} and unitary
\cite{Bender:2008gh,2000hep.th....1115M}. These analyses rely on a heuristic
comparison of CG to the Pais-Uhlenbech oscillator \cite{Pais:1950za}, but do
not provide a definite treatment of CG.

Since the effects of CG do not manifest themselves at short distances
\cite{Mannheim:2007ug}, we seek in this paper to understand what impact the
large-scale deviations will have on the morphology of a supermassive black
hole shadow. Section~\ref{sect:conformal} provides a short review of the
general CG formalism. In Section~\ref{sect:rotating}, we start with a brief
review of the CG Kerr metric for a rotating black hole and then we describe
the procedure used to obtain the BH shadows in conformal gravity. In
Section~\ref{sect:specific_cases}, we consider specific cases of BH shadows,
comparing GR and CG results; finally, in Section~\ref{sect:conclusions}, we
present our conclusions.


\section{\label{sect:conformal}Conformal gravity: a brief review}

Conformal gravity is based on the Weyl action:%

\begin{equation}
I_{W}=-\alpha_{g}\int d^{4}x\ (-g)^{1/2}\ C_{\lambda\mu\nu\kappa}%
\ C^{\lambda\mu\nu\kappa}, \label{eqn1.1}%
\end{equation}
where $g\equiv\det(g_{\mu\nu})$, $C_{\lambda\mu\nu\kappa}$ is the conformal
(or Weyl) tensor, and $\alpha_{g}$ is the CG coupling constant. $I_{W}$ is the
unique general coordinate scalar action that is invariant under local
conformal transformations: $g_{\mu\nu}(x)\rightarrow e^{2\alpha(x)}g_{\mu\nu
}(x)=\Omega^{2}(x)g_{\mu\nu}(x)$. The factor $\Omega(x)=e^{\alpha(x)}$
determines the amount of local stretching\ of the geometry, hence the name
conformal\ for a theory invariant under all local stretchings of the
space-time (see \cite{Varieschi:2008fc} and references therein for more details).

This conformally invariant generalization of GR was found to be a fourth-order
theory, as opposed to the standard second-order General Relativity, since the
field equations contained derivatives up to the fourth order of the metric
with respect to the space-time coordinates \cite{Bach:1921}. The fourth-order
CG field equations, $4\alpha_{g}W_{\mu\nu}=T_{\mu\nu}$ (where $W_{\mu\nu}$ is
the Bach tensor---see \cite{Mannheim:2005bfa,Varieschi:2008fc} for full
details) were studied in 1984 by Riegert \cite{Riegert:1984zz}, who obtained
the most general, spherically symmetric, static electrovacuum solution. The
explicit form of this solution, for the practical case of a static,
spherically symmetric source in CG, i.e., the fourth-order analogue of the
Schwarzschild exterior solution in GR, was then derived by Mannheim and
Kazanas in 1989 \cite{Mannheim:1988dj,Kazanas:1988qa}. This latter solution,
in the case $T_{\mu\nu}=0$ (exterior solution), is described by the metric%
\begin{equation}
ds^{2}=-B(r)\ c^{2}dt^{2}+\frac{dr^{2}}{B(r)}+r^{2}(d\theta^{2}+\sin^{2}%
\theta\ d\phi^{2}), \label{eqn1.2}%
\end{equation}
with%
\begin{equation}
B(r)=1-3\beta\gamma-\frac{\beta(2-3\beta\gamma)}{r}+\gamma r-\kappa r^{2}.
\label{eqn1.3}%
\end{equation}

The three integration constants in the last equation\ are as follows:
$\beta\ (%
\operatorname{cm}%
)$ can be considered the CG\ equivalent of the geometrized mass $\frac
{GM}{c^{2}}$, where $M$ is the mass of the (spherically symmetric) source and
$G$ is the universal gravitational constant; two additional parameters,
$\gamma\ (%
\operatorname{cm}%
^{-1})$ and $\kappa\ (%
\operatorname{cm}%
^{-2})$, are required by CG, while the standard Schwarzschild solution is
recovered for $\gamma,\kappa\rightarrow0$ in the equations above. The
quadratic term $-\kappa r^{2}$ indicates a background De Sitter spacetime,
which is important only over cosmological distances, since $\kappa$ has a very
small value. Similarly, $\gamma$ measures the departure from the Schwarzschild
metric at smaller distances, since the $\gamma r$ term becomes significant
over galactic distance scales.

The values of the CG parameters were first determined by Mannheim
\cite{Mannheim:2005bfa}:%
\begin{equation}
\gamma=3.06\times10^{-30}%
\operatorname{cm}%
^{-1},\ \kappa=9.54\times10^{-54}%
\operatorname{cm}%
^{-2} \label{eqn1.4}%
\end{equation}
and were also evaluated by one of us \cite{Varieschi:2008fc,Varieschi:2008va}
with a different approach, obtaining values which differ by a few orders of
magnitude from those above:%
\begin{equation}
\gamma=1.94\times10^{-28}%
\operatorname{cm}%
^{-1},\ \kappa=6.42\times10^{-48}%
\operatorname{cm}%
^{-2}. \label{eqn1.5}%
\end{equation}

Mannheim et al.
\cite{Mannheim:2005bfa,Mannheim:1992vj,Mannheim:1996rv,Mannheim:2010ti,Mannheim:2010xw,2012MNRAS.421.1273O,Mannheim:2012qw}
used the CG solutions in Eqs. (\ref{eqn1.2})-(\ref{eqn1.3}) to perform
extensive data fitting of galactic rotation curves without any dark matter
contribution, using the values of $\gamma$ and $\kappa$ as in Eq.
(\ref{eqn1.4}). Although the values of these CG\ parameters are very small,
the linear and quadratic terms in Eq. (\ref{eqn1.3}) become significant over
galactic and/or cosmological distances.

There has been some debate as to nature of CG's short distance (GR) and weak
field (Newtonian) limits. It was originally shown by Mannheim and Kazanas that
fourth-order conformal gravity respectively recovers both frameworks in the
appropriate limit \cite{Mannheim:1988dj,Mannheim:1992tr}. Although an
objection was raised by Flanagan in \cite{Flanagan:2006ra} that this may be
spoiled by the presence of a macroscopic scalar field that contributes to the
stress-energy tensor, it was later shown that this would still not influence
the Schwarzschild limit \cite{Mannheim:2007ug}. An additional criticism was
leveled by Yoon in \cite{Yoon:2013rxa} that the potential does not possess a
Newtonian limit
for classical point particles. It was shown very recently by Mannheim,
however, that this assumption violates the conformal invariance of the theory
and is an incorrect conclusion \cite{Mannheim:2015gba}.

\section{\label{sect:rotating}Rotating black holes in conformal gravity}

The standard GR Kerr metric is \cite{1992mtbh.book.....C} ($c=G=1$ in the following):%

\begin{equation}
ds^{2}=-\rho^{2}\frac{\Delta}{\Sigma^{2}}\ dt^{2}+\frac{\Sigma^{2}}{\rho^{2}%
}\left[  d\phi-\frac{2aMr}{\Sigma^{2}}dt\right]  ^{2}\sin^{2}\theta
\ +\frac{\rho^{2}}{\Delta}\ dr^{2}+\rho^{2}d\theta^{2}, \label{eqn2.1}%
\end{equation}
with%
\begin{equation}
\rho^{2}\equiv r^{2}+a^{2}\cos^{2}\theta\ ;\ \Delta\equiv r^{2}-2Mr+a^{2}%
\ ;\ \Sigma^{2}\equiv\left(  r^{2}+a^{2}\right)  ^{2}-a^{2}\Delta\sin
^{2}\theta. \label{eqn2.2}%
\end{equation}
In these equations $a$ is the angular momentum parameter ($a=J/M$) and $M$ is
the geometrized mass.

\subsection{\label{subsect:Kerr_metric}Kerr metric in CG}

The CG fourth-order Kerr metric, originally introduced by Mannheim and Kazanas
in 1991 \cite{Mannheim:1990ya}, can also be written in a similar
way\footnote{In general, CG\ quantities will be denoted by a tilde
(\symbol{126}) superscript.} \cite{Varieschi:2014ata}:%
\begin{equation}
ds^{2}=-\rho^{2}\frac{\widetilde{\Delta}_{r}\ \widetilde{\Delta}_{\theta}%
}{\widetilde{\Sigma}^{2}}\ dt^{2}+\frac{\widetilde{\Sigma}^{2}}{\rho^{2}%
}\left[  d\phi+\frac{\widetilde{\Delta}_{r}-\left(  r^{2}+a^{2}\right)
\widetilde{\Delta}_{\theta}}{\widetilde{\Sigma}^{2}}a\ dt\right]  ^{2}\sin
^{2}\theta\ +\frac{\rho^{2}}{\widetilde{\Delta}_{r}}\ dr^{2}+\frac{\rho^{2}%
}{\widetilde{\Delta}_{\theta}}d\theta^{2}, \label{eqn2.3}%
\end{equation}
with extended definitions for the CG auxiliary quantities and functions:%
\begin{align}
\widetilde{M}  &  \equiv M\left(  1-\frac{3}{2}\ M\gamma\right)
\ ;\ \widetilde{\Delta}_{r}\equiv r^{2}-2\widetilde{M}r+a^{2}-kr^{4}%
;\label{eqn2.4}\\
\widetilde{\Delta}_{\theta}  &  \equiv1-ka^{2}\cos^{2}\theta\cot^{2}%
\theta\ ;\ \widetilde{\Sigma}^{2}\equiv\widetilde{\Delta}_{\theta}\left(
r^{2}+a^{2}\right)  ^{2}-a^{2}\widetilde{\Delta}_{r}\sin^{2}\theta.\nonumber
\end{align}
We note that the CG fourth-order Kerr metric is conformal to the standard
second-order Kerr-de Sitter metric, as originally proven by Mannheim and
Kazanas \cite{Mannheim:1990ya}.

The constant $k$ is related to both parameters $\gamma$ and $\kappa$ of
conformal gravity:%

\begin{equation}
k=\kappa+\frac{\gamma^{2}(1-M\gamma)}{(2-3M\gamma)^{2}}. \label{eqn2.5}%
\end{equation}
However, these parameters $\gamma$ and $\kappa$ are very small; current
estimates give \cite{Mannheim:2005bfa,Varieschi:2008fc,Varieschi:2014pca}:%

\begin{align}
\gamma &  \sim10^{-30}-10^{-28}%
\operatorname{cm}%
^{-1},\label{eqn2.6}\\
\kappa &  \sim10^{-54}-10^{-48}%
\operatorname{cm}%
^{-2},\nonumber
\end{align}
as shown also in Eqs. (\ref{eqn1.4})-(\ref{eqn1.5}). It is easy to check that
for $\gamma,\kappa\rightarrow0$ the Kerr CG metric reduces to the standard
Kerr GR metric.

Dimensionless parameters to be used in the analysis of BH\ shadows are defined as:%

\begin{align}
a_{\ast}  &  =\frac{a}{M};\ \ \widetilde{a}_{\ast}=\frac{a}{\widetilde{M}%
};\label{eqn2.7}\\
\gamma_{\ast}  &  =\gamma M\sim10^{-24}-10^{-14};\nonumber\\
k_{\ast}  &  =k\widetilde{M}^{2}\sim10^{-42}-10^{-20}.\nonumber
\end{align}
The condition for the existence of the GR event horizon is $\left\vert
a_{\ast}\right\vert \leq1$, obtained from the equation $\Delta=r^{2}%
-2Mr+a^{2}=0$; in CG\ the equivalent equation $\widetilde{\Delta}_{r}%
=r^{2}-2\widetilde{M}r+a^{2}-kr^{4}=0$ will be solved numerically, for the
cases of interest in Sect. \ref{sect:specific_cases}.

The ranges for $\gamma_{\ast}$ and $k_{\ast}$ in Eq. (\ref{eqn2.7}) were
estimated as follows: we assume BH\ masses in the range $\widetilde{M}\sim
M\sim\left(  10-10^{9}\right)  M_{\odot}\sim10^{6}-10^{14}%
\operatorname{cm}%
$, since BH masses range \cite{Li:2013jra} from $M\sim10M_{\odot}$ (dark
compact objects) to $M\sim10^{9}M_{\odot}$ (heaviest super-massive BH), with
$M_{\odot}\sim10^{5}%
\operatorname{cm}%
$. These numbers are then combined with the ranges for $\gamma$ and $\kappa$
in Eq. (\ref{eqn2.6}) and with $k\approx\kappa$, due to Eq. (\ref{eqn2.5}).

Following \cite{Li:2013jra}, \cite{1992mtbh.book.....C}, and
\cite{Varieschi:2014ata} the radial equation of motion for photons is:%

\begin{equation}
\overset{\centerdot}{r}^{2}=\left(  \frac{dr}{d\tau}\right)  ^{2}=%
\begin{Bmatrix}
\frac{\widetilde{R}(r)}{\rho^{4}}=\frac{\widetilde{\Delta}_{r}{}^{2}}{\rho
^{4}}p_{r}^{2}\ ;\text{ 4th-order CG}\\
\frac{R(r)}{\rho^{4}}=\frac{\Delta^{2}}{\rho^{4}}p_{r}^{2}\ ;\text{ 2nd-order
GR}%
\end{Bmatrix}
, \label{eqn2.8}%
\end{equation}
with $\tau$ being an affine parameter, and (for photons)%

\begin{align}
\widetilde{R}(r)  &  \equiv\left[  \left(  r^{2}+a^{2}\right)  E-aL_{z}%
\right]  ^{2}-\widetilde{\Delta}_{r}[\widetilde{\mathcal{Q}}+(L_{z}%
-aE)^{2}]\label{eqn2.9}\\
&  =[E^{2}+k((aE-L_{z})^{2}+\widetilde{\mathcal{Q}})]r^{4}+(a^{2}E^{2}%
-L_{z}^{2}-\widetilde{\mathcal{Q}}\mathcal{)}r^{2}+2\widetilde{M}%
[(aE-L_{z})^{2}+\widetilde{\mathcal{Q}}\mathcal{]}r-a^{2}%
\widetilde{\mathcal{Q}}\nonumber\\
R(r)  &  \equiv\left[  \left(  r^{2}+a^{2}\right)  E-aL_{z}\right]
^{2}-\Delta\lbrack\mathcal{Q+}(L_{z}-aE)^{2}]\nonumber\\
&  =E^{2}r^{4}+(a^{2}E^{2}-L_{z}^{2}-\mathcal{Q)}r^{2}+2M[(aE-L_{z}%
)^{2}+\mathcal{Q]}r-a^{2}\mathcal{Q},\nonumber
\end{align}
respectively, in the CG and GR cases. $E$ and $L_{z}$ are interpreted
respectively as energy per unit mass and angular momentum---in the axial
direction---per unit mass, while Carter's constant (for photons) is also
different in the two cases:%

\begin{align}
\widetilde{\mathcal{Q}}  &  =\widetilde{\Delta}_{\theta}p_{\theta}^{2}%
+\frac{(aE\sin\theta-L_{z}\csc\theta)^{2}}{\widetilde{\Delta}_{\theta}}%
-(L_{z}-aE)^{2}\label{eqn2.10}\\
\mathcal{Q}  &  =p_{\theta}^{2}+(aE\sin\theta-L_{z}\csc\theta)^{2}%
-(L_{z}-aE)^{2}\nonumber\\
&  =p_{\theta}^{2}+\cos^{2}\theta\left(  \frac{L_{z}^{2}}{\sin^{2}\theta
}-a^{2}E^{2}\right)  .\nonumber
\end{align}
We note that Carter's constant in CG ($\widetilde{\mathcal{Q}}$) cannot be
written in a form similar to the one for $\mathcal{Q}$\ in the last line of
the previous equation.

It is customary to minimize the parameters by setting:%
\begin{equation}
\xi=\widetilde{\xi}=L_{z}/E\ ;\ \eta=\mathcal{Q}/E^{2}\ ;\ \widetilde{\eta
}=\widetilde{\mathcal{Q}}/E^{2} \label{eqn2.11}%
\end{equation}
and rewrite Eq. (\ref{eqn2.9}) in terms of modified functions (rescaled
functions, dividing by $E^{2}$): $\mathit{R}(r)=R(r)/E^{2}$ and
$\widetilde{\mathit{R}}(r)=\widetilde{R}(r)/E^{2}$. These modified radial
functions (for photons) can be written explicitly as:%
\begin{align}
\widetilde{\mathit{R}}(r)  &  =[1+k(\widetilde{\eta}+(\widetilde{\xi}%
-a)^{2})]r^{4}+(a^{2}-\widetilde{\xi}^{2}-\widetilde{\eta})r^{2}%
+2\widetilde{M}[\widetilde{\eta}+(\widetilde{\xi}-a)^{2}]r-a^{2}%
\widetilde{\eta}\label{eqn2.12}\\
\mathit{R}(r)  &  =r^{4}+(a^{2}-\xi^{2}-\eta)r^{2}+2M[\eta+(\xi-a)^{2}%
]r-a^{2}\eta\nonumber
\end{align}
where, as usual, the CG case reduces to the GR case for $\gamma,\kappa
\rightarrow0$.

Similarly, rescaled angular functions $\mathit{\Theta}(\theta)=\Theta
(\theta)/E^{2}$ and $\widetilde{\mathit{\Theta}}(\theta)=\widetilde{\Theta
}(\theta)/E^{2}$ are also introduced (see \cite{1992mtbh.book.....C} and
\cite{Varieschi:2014ata} for details):%

\begin{align}
\widetilde{\mathit{\Theta}}(\theta)  &  =\widetilde{\Delta}_{\theta
}\widetilde{\eta}+a^{2}\cos^{2}\theta+(\widetilde{\Delta}_{\theta
}-1)(\widetilde{\xi}-a)^{2}-\widetilde{\xi}^{2}\cot^{2}\theta\label{eqn2.13}\\
\mathit{\Theta}(\theta)  &  =\eta+a^{2}\cos^{2}\theta-\xi^{2}\cot^{2}%
\theta\nonumber
\end{align}

\subsection{\label{subsect:BH_shadows}Black hole shadows in CG}

BH\ shadows are related to unstable circular photon orbits. These are obtained
by setting $\widetilde{\mathit{R}}(r)=\frac{\partial\widetilde{\mathit{R}}%
}{\partial r}=0$ and $\mathit{R}(r)=\frac{\partial\mathit{R}}{\partial r}=0$
for our two cases. Explicitly:%

\begin{align}
\lbrack1+k(\widetilde{\eta}_{c}+(\widetilde{\xi}_{c}-a)^{2})]r^{4}%
+(a^{2}-\widetilde{\xi}_{c}^{2}-\widetilde{\eta}_{c})r^{2}+2\widetilde{M}%
[\widetilde{\eta}_{c}+(\widetilde{\xi}_{c}-a)^{2}]r-a^{2}\widetilde{\eta}_{c}
&  =0\label{eqn2.14}\\
4[1+k(\widetilde{\eta}_{c}+(\widetilde{\xi}_{c}-a)^{2})]r^{3}+2(a^{2}%
-\widetilde{\xi}_{c}^{2}-\widetilde{\eta}_{c})r+2\widetilde{M}[\widetilde{\eta
}_{c}+(\widetilde{\xi}_{c}-a)^{2}]  &  =0\nonumber
\end{align}
and%

\begin{align}
r^{4}+(a^{2}-\xi_{c}^{2}-\eta_{c})r^{2}+2M[\eta_{c}+(\xi_{c}-a)^{2}%
]r-a^{2}\eta_{c}  &  =0\label{eqn2.15}\\
4r^{3}+2(a^{2}-\xi_{c}^{2}-\eta_{c})r+2M[\eta_{c}+(\xi_{c}-a)^{2}]  &
=0\nonumber
\end{align}
where $(\widetilde{\xi}_{c},\widetilde{\eta}_{c})$ and $(\xi_{c},\eta_{c})$
represent the critical loci, i.e., the set of unstable circular photon orbits
in the two cases.

Following the procedure outlined in \cite{1992mtbh.book.....C} or in
\cite{2000CQGra..17..123D}, we can solve these equations for the CG case and obtain:%

\begin{align}
\widetilde{\xi}_{c}  &  =\frac{[\widetilde{M}(r^{2}-a^{2})-r(r^{2}%
-2\widetilde{M}r+a^{2})-2a^{2}kr^{3}]}{a(r-\widetilde{M}-2kr^{3})}%
=\frac{[\widetilde{M}(r^{2}-a^{2})-r\widetilde{\Delta}_{r}-kr^{3}(r^{2}%
+2a^{2})]}{a(r-\widetilde{M}-2kr^{3})}\label{eqn2.16}\\
\widetilde{\eta}_{c}  &  =\frac{r^{3}[4a^{2}\widetilde{M}-r(r-3\widetilde{M}%
)^{2}-4a^{2}kr^{3}]}{a^{2}(r-\widetilde{M}-2kr^{3})^{2}}=\frac{r^{3}%
[4\widetilde{M}\widetilde{\Delta}_{r}+4kr^{3}(\widetilde{M}r-a^{2}%
)-r(r-\widetilde{M})^{2}]}{a^{2}(r-\widetilde{M}-2kr^{3})^{2}},\nonumber
\end{align}
similar to the second-order solutions:%

\begin{align}
\xi_{c}  &  =\frac{[M(r^{2}-a^{2})-r(r^{2}-2Mr+a^{2})]}{a(r-M)}=\frac
{[M(r^{2}-a^{2})-r\Delta]}{a(r-M)}\label{eqn2.17}\\
\eta_{c}  &  =\frac{r^{3}[4a^{2}M-r(r-3M)^{2}]}{a^{2}(r-M)^{2}}=\frac
{r^{3}[4M\Delta-r(r-M)^{2}]}{a^{2}(r-M)^{2}}.\nonumber
\end{align}

Since plots of BH\ shadows are usually done with coordinates expressed in
units of mass, it is more practical to rewrite the previous solutions in Eqs.
(\ref{eqn2.16}) and (\ref{eqn2.17}) in terms of dimensionless quantities
(capital Xi and capital Eta) as follows:%

\begin{align}
\widetilde{\Xi}_{c}  &  =\frac{\widetilde{\xi}_{c}}{\widetilde{M}}%
=\frac{[(\widetilde{z}^{2}-\widetilde{a}_{\ast}^{2})-\widetilde{z}%
(\widetilde{z}^{2}-2\widetilde{z}+\widetilde{a}_{\ast}^{2})-2\widetilde{a}%
_{\ast}^{2}k_{\ast}\widetilde{z}^{3}]}{\widetilde{a}_{\ast}(\widetilde{z}%
-1-2k_{\ast}\widetilde{z}^{3})}\label{eqn2.18}\\
\widetilde{H}_{c}  &  =\frac{\widetilde{\eta}_{c}}{\widetilde{M}^{2}}%
=\frac{\widetilde{z}^{3}[4\widetilde{a}_{\ast}^{2}-\widetilde{z}%
(\widetilde{z}-3)^{2}-4\widetilde{a}_{\ast}^{2}k_{\ast}\widetilde{z}^{3}%
]}{\widetilde{a}_{\ast}^{2}(\widetilde{z}-1-2k_{\ast}\widetilde{z}^{3})^{2}%
},\nonumber
\end{align}
and%

\begin{align}
\Xi_{c}  &  =\frac{\xi_{c}}{M}=\frac{[(z^{2}-a_{\ast}^{2})-z(z^{2}-2z+a_{\ast
}^{2})]}{a_{\ast}(z-1)}\label{eqn2.19}\\
H_{c}  &  =\frac{\eta_{c}}{M^{2}}=\frac{z^{3}[4a_{\ast}^{2}-z(z-3)^{2}%
]}{a_{\ast}^{2}(z-1)^{2}}.\nonumber
\end{align}
In the previous two equations, we used the dimensionless radial variables
$z=r/M$ and $\widetilde{z}=r/\widetilde{M}$, and also the dimensionless
parameters from Eq. (\ref{eqn2.7}).

The last step in the procedure for plotting BH\ shadows is to consider the
celestial coordinates $x$ and $y$ of the image, as seen by an observer at
infinity. The standard GR\ procedure
\cite{1992mtbh.book.....C,2000CQGra..17..123D}, considers the tetrad
components of the four momentum as:%

\begin{align}
p^{(t)}  &  =e^{-\nu}(E-\omega L_{z})=\frac{\Sigma}{\rho\sqrt{\Delta}}\left(
E-\frac{2aMr}{\Sigma^{2}}L_{z}\right) \label{eqn2.20}\\
p^{(r)}  &  =-e^{-\mu_{2}}p_{r}=\frac{\sqrt{\Delta}}{\rho}\frac{\sqrt{R(r)}%
}{\Delta}=\frac{1}{\rho}\sqrt{\frac{R(r)}{\Delta}}\nonumber\\
p^{(\theta)}  &  =-e^{-\mu_{3}}p_{\theta}=\frac{1}{\rho}\sqrt{\Theta(\theta
)}=\frac{1}{\rho}\sqrt{\left(  \eta+a^{2}\cos^{2}\theta-\xi^{2}\cot^{2}%
\theta\right)  E^{2}}\nonumber\\
p^{(\phi)}  &  =e^{-\psi}L_{z}=\frac{\rho}{\sin\theta\ \Sigma}L_{z},\nonumber
\end{align}
where $\nu$, $\omega$, $\mu_{2}$, $\mu_{3}$, and $\psi$ are functions of $r$
and $\theta$, which can be expressed in terms of all the other functions used previously.

The celestial coordinates $x$ and $y$ of the image are then computed in terms
of $\xi$, $\eta$, and of the angular coordinate of the observer at infinity,
$\theta\rightarrow i$:%

\begin{align}
x  &  =\left(  \frac{rp^{(\phi)}}{p^{(t)}}\right)  _{r\rightarrow\infty}%
=\frac{\xi}{\sin i}\label{eqn2.21}\\
y  &  =\left(  \frac{rp^{(\theta)}}{p^{(t)}}\right)  _{r\rightarrow\infty}%
=\pm\left(  \eta+a^{2}\cos^{2}i-\xi^{2}\cot^{2}i\right)  ^{1/2},\nonumber
\end{align}
where the previous equation is obtained by using the quantities in Eq.
(\ref{eqn2.20}) and taking limits for $r\rightarrow\infty$.

Rescaling also the celestial coordinates into dimensionless ones, $X=x/M$ and
$Y=y/M$, and combining together Eqs. (\ref{eqn2.19}) and (\ref{eqn2.21}),
yields the parametric form of the critical locus in dimensionless coordinates:%

\begin{align}
X  &  =\frac{x}{M}=\frac{\xi_{c}/M}{\sin i}=\frac{\Xi_{c}}{\sin i}=\frac
{1}{\sin i}\frac{[(z^{2}-a_{\ast}^{2})-z(z^{2}-2z+a_{\ast}^{2})]}{a_{\ast
}(z-1)}\label{eqn2.22}\\
Y  &  =\frac{y}{M}=\pm\left(  \frac{\eta_{c}}{M^{2}}+\frac{a^{2}}{M^{2}}%
\cos^{2}i-\frac{\xi_{c}^{2}}{M^{2}}\cot^{2}i\right)  ^{1/2}=\pm\left(
H_{c}+a_{\ast}^{2}\cos^{2}i-\Xi_{c}^{2}\cot^{2}i\right)  ^{1/2}\nonumber\\
&  =\pm\left\{  \frac{z^{3}[4a_{\ast}^{2}-z(z-3)^{2}]}{a_{\ast}^{2}(z-1)^{2}%
}+a_{\ast}^{2}\cos^{2}i-\left[  \frac{(z^{2}-a_{\ast}^{2})-z(z^{2}-2z+a_{\ast
}^{2})}{a_{\ast}(z-1)}\right]  ^{2}\cot^{2}i\right\}  ^{1/2}.\nonumber
\end{align}
This locus can be plotted for values of the parameter $z\gtrsim1+\sqrt
{1-a_{\ast}^{2}}$ obtaining the standard GR shadows that will be shown in
Sect. \ref{sect:specific_cases}.

The same procedure can be repeated in CG; the equivalent of Eq. (\ref{eqn2.20}%
) in CG is:%

\begin{align}
p^{(t)}  &  =e^{-\nu}(E-\omega L_{z})=\frac{\widetilde{\Sigma}}{\rho
\sqrt{\widetilde{\Delta}_{r}\widetilde{\Delta}_{\theta}}}\left(
E-\frac{-\widetilde{\Delta}_{r}+(r^{2}+a^{2})\widetilde{\Delta}_{\theta}%
}{\widetilde{\Sigma}^{2}}aL_{z}\right) \label{eqn2.23}\\
p^{(r)}  &  =-e^{-\mu_{2}}p_{r}=\frac{\sqrt{\widetilde{\Delta}_{r}}}{\rho
}\frac{\sqrt{\widetilde{R}(r)}}{\widetilde{\Delta}_{r}}=\frac{1}{\rho}%
\sqrt{\frac{\widetilde{R}(r)}{\widetilde{\Delta}_{r}}}\nonumber\\
p^{(\theta)}  &  =-e^{-\mu_{3}}p_{\theta}=\frac{\sqrt{\widetilde{\Delta
}_{\theta}}}{\rho}\frac{\sqrt{\widetilde{\Theta}(\theta)}}{\widetilde{\Delta
}_{\theta}}=\frac{1}{\rho}\sqrt{\frac{\widetilde{\Delta}_{\theta
}\widetilde{\eta}+a^{2}\cos^{2}\theta+(\widetilde{\Delta}_{\theta
}-1)(\widetilde{\xi}-a)^{2}-\widetilde{\xi}^{2}\cot^{2}\theta}%
{\widetilde{\Delta}_{\theta}}E^{2}}\nonumber\\
p^{(\phi)}  &  =e^{-\psi}L_{z}=\frac{\rho}{\sin\theta\ \widetilde{\Sigma}%
}L_{z}.\nonumber
\end{align}

The celestial coordinates $\widetilde{x}$ and $\widetilde{y}$ of the image, as
a function of $\widetilde{\xi}$, $\widetilde{\eta}$, and of the angular
coordinate of the observer at infinity, $\theta\rightarrow i$, are obtained by
considering the leading terms for large $r$ ($r\gg a$ and $r\gg\widetilde{M}%
$). CG terms proportional to $ka$ are neglected, due to the small value of
$k$, but terms such as $\left(  1-kr^{2}\right)  ^{1/2}$ and $\left(
1-ka^{2}\cos^{2}i\cot^{2}i\right)  ^{1/2}$\ are included, as possible CG corrections:%

\begin{align}
\widetilde{x}  &  =\left(  \frac{rp^{(\phi)}}{p^{(t)}}\right)  _{r\gg
a,\widetilde{M}}=\frac{\widetilde{\xi}}{\sin i}\frac{\left(  1-kr^{2}\right)
^{1/2}}{\left(  1-ka^{2}\cos^{2}i\cot^{2}i\right)  ^{1/2}}\label{eqn2.24}\\
\widetilde{y}  &  =\left(  \frac{rp^{(\theta)}}{p^{(t)}}\right)  _{r\gg
a,\widetilde{M}}=\pm\left[  \left(  1-ka^{2}\cos^{2}i\cot^{2}i\right)
\widetilde{\eta}+a^{2}\cos^{2}i-ka^{2}\cos^{2}i\cot^{2}i(\widetilde{\xi
}-a)^{2}-\widetilde{\xi}^{2}\cot^{2}i\right]  ^{1/2}\nonumber\\
&  \times\frac{\left(  1-kr^{2}\right)  ^{1/2}}{\left(  1-ka^{2}\cos^{2}%
i\cot^{2}i\right)  ^{1/2}}.\nonumber
\end{align}
Due to the presence of the $\left(  1-kr^{2}\right)  ^{1/2}$ terms, the
CG\ expressions for $\widetilde{x}$ and $\widetilde{y}$ are divergent for
$r\rightarrow\infty$, but they can be considered as valid approximations for
$r<\sqrt{k^{-1}}\sim10^{24}-10^{27}%
\operatorname{cm}%
$.

The dimensionless celestial coordinates $\widetilde{X}=\widetilde{x}%
/\widetilde{M}$ and $\widetilde{Y}=\widetilde{y}/\widetilde{M}$ in CG become:%

\begin{align}
\widetilde{X}  &  =\frac{\widetilde{x}}{\widetilde{M}}=\frac{\widetilde{\Xi
}_{c}}{\sin i}\frac{\left(  1-k_{\ast}\widetilde{z}^{2}\right)  ^{1/2}%
}{\left(  1-k_{\ast}\widetilde{a}_{\ast}^{2}\cos^{2}i\cot^{2}i\right)  ^{1/2}%
}\label{eqn2.25}\\
\widetilde{Y}  &  =\frac{\widetilde{y}}{\widetilde{M}}=\pm\left[  (1-k_{\ast
}\widetilde{a}_{\ast}^{2}\cos^{2}i\cot^{2}i)\widetilde{H}_{c}+\widetilde{a}%
_{\ast}^{2}\cos^{2}i-k_{\ast}\widetilde{a}_{\ast}^{2}\cos^{2}i\cot
^{2}i\ (\widetilde{\Xi}_{c}-\widetilde{a}_{\ast})^{2}-\widetilde{\Xi}_{c}%
^{2}\cot^{2}i\right]  ^{1/2}\nonumber\\
&  \times\frac{\left(  1-k_{\ast}\widetilde{z}^{2}\right)  ^{1/2}}{\left(
1-k_{\ast}\widetilde{a}_{\ast}^{2}\cos^{2}i\cot^{2}i\right)  ^{1/2}},\nonumber
\end{align}
which correctly reduce to the GR\ dimensionless coordinates in Eq.
(\ref{eqn2.22}) for $k_{\ast}\rightarrow0$ and $\widetilde{M}\rightarrow M$.
In the next section, we will use the GR and CG expressions for the
dimensionless celestial coordinates (equations (\ref{eqn2.22}) and
(\ref{eqn2.25}), respectively), in order to plot the black hole shadows in
these two cases.

Since the GR and CG shadow plots will be directly compared with each other, we
will need all coordinates to be expressed in terms of the standard mass $M$,
i.e., the $\widetilde{X}$ and $\widetilde{Y}$ coordinates need to be rescaled
by a common factor $(1-\frac{3}{2}M\gamma)=(1-\frac{3}{2}\gamma_{\ast})$, in
view of Eqs. (\ref{eqn2.4}) and (\ref{eqn2.7}). Explicitly:%

\begin{align}
X_{(CG)}  &  =\frac{\widetilde{x}}{M}=\widetilde{X}\frac{\widetilde{M}}%
{M}=\widetilde{X}\left(  1-\frac{3}{2}\gamma_{\ast}\right) \label{eqn2.26}\\
Y_{(CG)}  &  =\frac{\widetilde{y}}{M}=\widetilde{Y}\frac{\widetilde{M}}%
{M}=\widetilde{Y}\left(  1-\frac{3}{2}\gamma_{\ast}\right)  ,\nonumber
\end{align}
where $X_{(CG)}$ and $Y_{(CG)}$ indicate these rescaled CG coordinates, to be
compared with the rescaled GR coordinates in Eq. (\ref{eqn2.22}).

\section{\label{sect:specific_cases}Specific cases of BH shadows}

In this section, we will plot the CG black hole shadows, following Eqs.
(\ref{eqn2.25})-(\ref{eqn2.26}), and compare them with the GR\ shadows,
following Eq. (\ref{eqn2.22}), for several cases of interest.

As our first example, we consider the case of Sagittarius A*, the supermassive
black hole at the center of our galaxy, already mentioned in Sect.
\ref{sect:introduction}. Its mass and distance are estimated as follows:%

\begin{align}
M  &  =%
\begin{Bmatrix}
(4.31\pm0.38)\times10^{6}M_{\odot}=\left(  6.36\pm0.56\right)  \times10^{11}%
\operatorname{cm}%
\ \text{\cite{Gillessen:2008qv}}\\
(4.1\pm0.6)\times10^{6}M_{\odot}=\left(  6.1\pm0.9\right)  \times10^{11}%
\operatorname{cm}%
\ \text{\cite{Ghez:2008ms}}\\
(4.02\pm0.16)\times10^{6}M_{\odot}=\left(  5.94\pm0.24\right)  \times10^{11}%
\operatorname{cm}%
\ \text{\cite{2016arXiv160705726B}}%
\end{Bmatrix}
\label{eqn3.1}\\
r  &  =%
\begin{Bmatrix}
(7,940\pm420)\ pc=(2.45\pm0.13)\times10^{22}%
\operatorname{cm}%
\ \text{\cite{Eisenhauer:2003he}}\\
(7,860\pm140)\ pc=(2.425\pm0.043)\times10^{22}%
\operatorname{cm}%
\ \text{\cite{2016arXiv160705726B}}%
\end{Bmatrix}
\nonumber
\end{align}

\begin{figure}[ptb]
\includegraphics[scale=0.7]{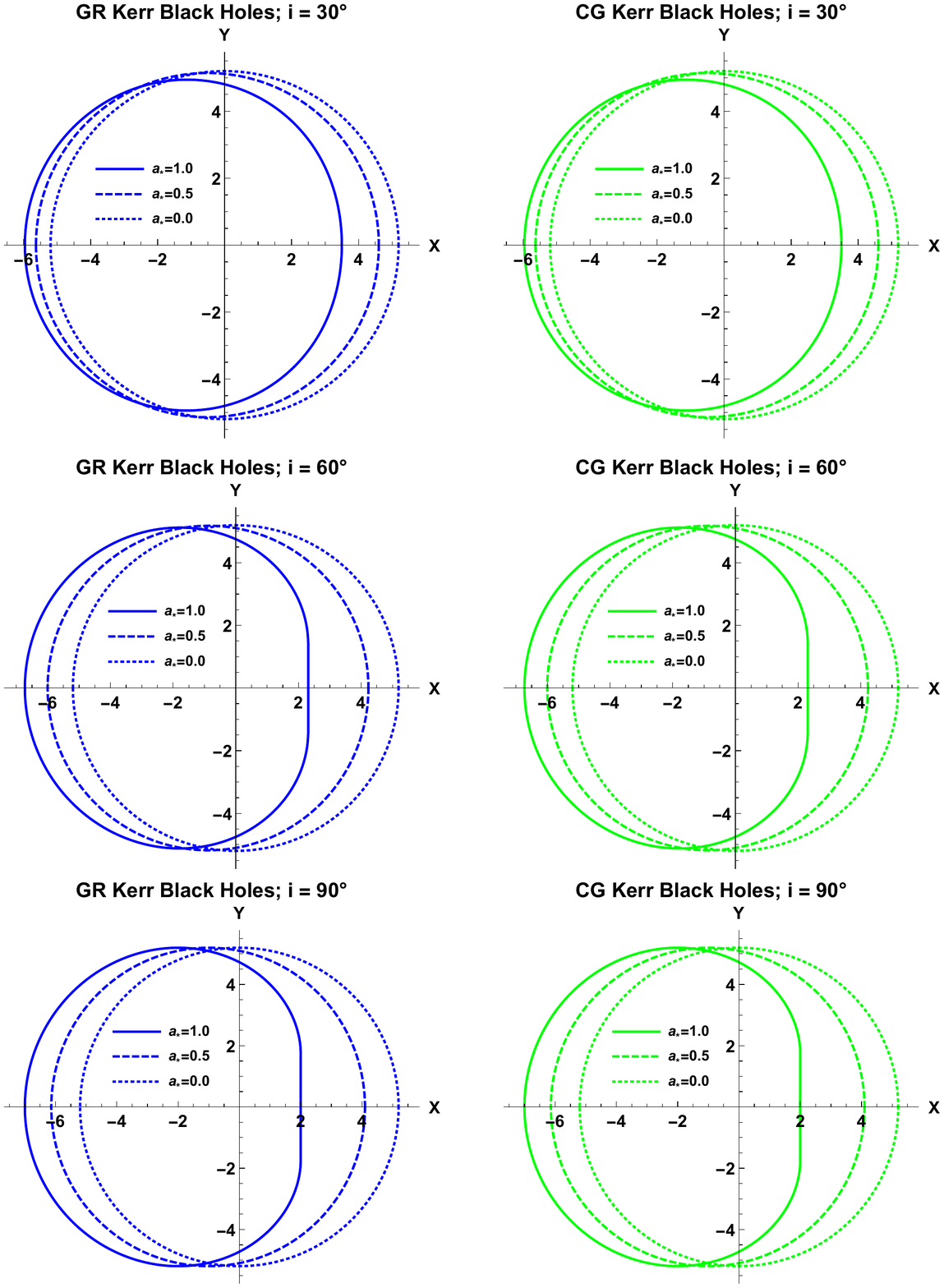}\caption{Kerr black hole shadows,
computed following GR and CG, for different values of the angular coordinate
$i$ and of the dimensionless parameter $a_{\ast}$. The GR shadows (left
panels, in blue) are independent of the black hole mass, while the CG shadows
(right panels, in green) are computed using the mass of Sgr A* black hole and
the CG parameters from Eq. (\ref{eqn1.5}). There are practically no
differences between GR and CG\ plots. }%
\label{fig1}%
\end{figure}

In Fig. 1 we plot the GR and CG\ shadows of Sgr A*, for different values of
the angular coordinate $i$ and of the dimensionless angular momentum parameter
$a_{\ast}$. The GR shadow plots are independent of the black hole mass, since
they use the dimensionless coordinates $X$ and $Y$ from Eq. (\ref{eqn2.22}).
The CG plots use instead the rescaled dimensionless coordinates $X_{(CG)}$ and
$Y_{(CG)}$ from Eqs. (\ref{eqn2.25})-(\ref{eqn2.26}), which depend on the
black hole mass through the factor $(1-\frac{3}{2}\gamma_{\ast})=(1-\frac
{3}{2}M\gamma)$.

The CG\ shadows also depend critically on the CG\ parameters $\gamma$ and
$\kappa$ (or the equivalent $\gamma_{\ast}$ and $k_{\ast}$); for the plots in
Fig. 1 we used the values shown in Eq. (\ref{eqn1.5}), which represent the
largest estimates of these parameters currently in the literature. However,
given their small values, no significant differences are noticeable in Fig. 1
between the GR\ shadows (left panels, in blue) and the CG shadows (right
panels, in green). The differences between the GR and CG\ plots for
Sagittarius A* were estimated to be on the order of $10^{-15}$, i.e., Kerr
black hole shadows computed in CG are virtually the same as those in GR.

Given that EHT observations of the Sgr A* shadow are likely to measure its
angular size as $R\simeq(26.4\pm$ $1.5)\ \mu\operatorname{arcsec}$
\cite{Johannsen:2015hib}, i.e., with a $\sim6\%$ uncertainty of the angular
radius, no practical differences are expected between GR and CG\ shadows for
Sagittarius A*, due to the extremely small CG corrections ($\sim10^{-15}$)
estimated above.

As our second case, we consider a supermassive black hole with $M\sim
10^{11}M_{\odot}\sim10^{16}%
\operatorname{cm}%
$, which corresponds to the largest current estimates of black hole masses,
such as the S5 0014+813 supermassive BH \cite{2009MNRAS.399L..24G} or similar,
and we also increase the values of the CG\ parameters $\gamma$ and $\kappa$,
in order to obtain significant differences between GR and CG shadows.

\begin{figure}[ptb]
\caption{ Kerr black hole shadows, computed following CG with $M=10^{16}%
\operatorname{cm}$, for different values of the angular coordinate $i$ and of
the dimensionless parameter $a_{\ast}$. The CG shadows in the left panels (in
red) are computed with $\kappa\approx10^{-34}\operatorname{cm}^{-2}$ , while
the CG shadows in the right panels (in cyan) are computed with $\gamma
\approx10^{-17}\operatorname{cm}^{-1}$. For these values of the CG parameters
there are noticeable differences between CG plots in this Fig. 2 and GR\ plots
from Fig. 1 }%
\label{fig2}
\includegraphics[scale=0.7]{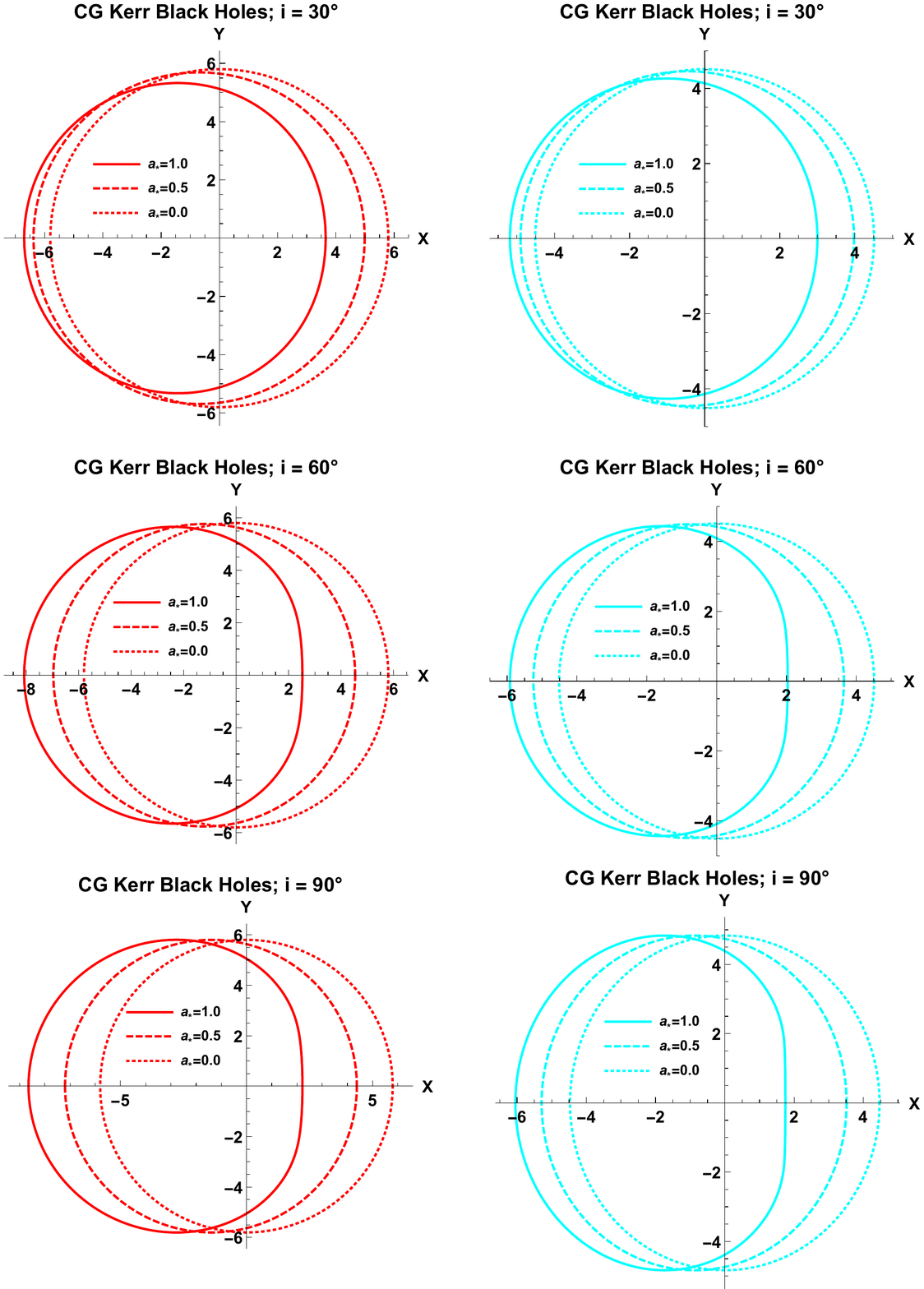}\end{figure}

The CG\ shadows in the left panels of Fig. 2 (in red) were obtained by using
$M=10^{16}%
\operatorname{cm}%
$, $\gamma=1.94\times10^{-28}%
\operatorname{cm}%
^{-1}$ (same value for $\gamma$ as in Eq. (\ref{eqn1.5})), and by increasing
the $\kappa$ parameter by several orders of magnitudes:$\ \kappa
\approx10^{-34}%
\operatorname{cm}%
^{-2}$. Similarly, the CG\ shadows in the right panels of Fig. 2 (in cyan)
were obtained by using $M=10^{16}%
\operatorname{cm}%
$, $\kappa=6.42\times10^{-48}%
\operatorname{cm}%
^{-2}$ (same value for $\kappa$ as in Eq. (\ref{eqn1.5})), and by increasing
the $\gamma$ parameter by several orders of magnitudes:$\ \gamma
\approx10^{-17}%
\operatorname{cm}%
^{-1}$.

All these CG\ shadows in Fig. 2 should be compared with the respective GR
shadows in Fig. 1 (which are mass independent). The comparison now shows
noticeable differences between GR and CG cases. However, the CG\ plots in Fig.
2 were obtained by using large values of the CG\ parameters $\gamma$ and
$\kappa$, which are currently ruled out by CG analysis of galactic rotation
curves and of the cosmological accelerated expansion of the Universe.

Therefore, we conclude that CG Kerr black hole shadows are not likely to look
any different from the equivalent GR shadows and thus the Event Horizon
Telescope will probably not be able to differentiate between predictions of
standard GR and of alternative CG theories.

\section{\label{sect:conclusions}Conclusion}

In this paper, we have analyzed the morphology of shadows from a rotating,
neutral black hole in a fourth-order conformal Weyl gravity framework, to
determine any potentially observable model-dependent characteristics. Since
Weyl gravitation provides large-scale modifications to general relativity, the
horizon size and mass of Sgr A* should provide a suitable testbed for this
theory. Unfortunately, we have shown that the shadow morphology is not
affected by the underlying framework, giving a deviation from the GR case on
the order of $10^{-15}$ in the dimensionless coordinates.

A difference measurable by the EHT (\textit{i.e.} greater than $6\%$ of the
angular radius) only arises for the largest known supermassive black holes on
the order of $M\sim10^{10}-10^{11}M_{\odot}$, well above the estimated mass
range of Sgr A* ($10^{6}M_{\odot}$). In this case, however, the shadow
characteristics would be measurably different from those predicted by general
relativity if the constraints on the conformal parameters $\gamma$ and
$\kappa$ are sufficiently loose and increased by several orders of magnitude.
This would push the parameters well outside the range provided by experimental
observations, however. It is thus unlikely that any differentiable shadow
characteristics from pure conformal Weyl gravity will be detected in upcoming experiments.

Since it is anticipated that quantum effects will become realizable on the
macroscopic horizon scales of such supermassive BHs, however, one could
consider extensions to Weyl CG that include quantum corrections. These might
include adding a minimal length scale \cite{Hossenfelder:2012jw}, a
non-commutative geometry \cite{Nicolini:2008aj}, the Generalized Uncertainty
Principle \cite{Carr:2015nqa}, or asymptotic safety \cite{Reuter:2005bb}. It
is possible that such a quantum/cosmological hybrid model could produce the
observable effects discussed herein, while bringing the CG parameters within
their experimentally constrained range. These quantum effects are currently
being investigated by the authors.


\bibliographystyle{apsrev}
\bibliography{BHshadowsv7}

\end{document}